\documentclass[%
 reprint,
 amsmath,amssymb,
 aps,
 pre,
floatfix,
longbibliography,
]{revtex4-2}

\usepackage{graphicx}
\usepackage{dcolumn}
\usepackage{bm}
\usepackage{siunitx}  
\usepackage[version=4]{mhchem}
\DeclareSIUnit{\torr}{Torr}
\usepackage{hyperref}
\usepackage{enumitem}

\begin{document}

\title{Aging in the Flow Dynamics of Dense Suspensions of Contactless Microparticles}

\author{Jesús Fernández}
\author{Loïc Vanel}
\author{Antoine Bérut}
\email{Corresponding author: antoine.berut@univ-lyon1.fr}
\affiliation{Universite Claude Bernard Lyon 1, CNRS, Institut Lumière Matière, UMR5306, F-69100, Villeurbanne, France}

\date{\today}

\begin{abstract}

This study demonstrates that the free-surface flow dynamics of dense piles of contactless silica microparticles depend on the resting period prior to flow. Microfluidic rotating drum experiments reveal that longer resting times lead to delayed flow onsets and reduced flow velocities, both evolving logarithmically with the resting time. These aging-like effects are more pronounced for thermally driven creep flows in piles with initial tilting angle below the athermal angle of repose, in contrast to piles initially tilted above this repose angle, where gravity-driven flows tend to gradually erase aging effects. Moreover, we show that the packing fraction does not change during the resting period, and that aging occurs in both monodisperse and polydisperse piles, indicating that crystallization is not required for the time-dependent behavior to appear. Remarkably, vigorous agitation that re-disperses the particles fully restores the piles to their initial state, demonstrating that the observed effects are not due to sample degradation. These findings evidence a form of aging in quiescent suspensions intermediate between colloidal and granular media, where thermal fluctuations, still significant relative to particle weight, progressively stabilize the system, making it more resistant to flow and deformation.

\end{abstract}

\maketitle


\section{INTRODUCTION}
\label{sec:introduction}

Understanding the evolution and stability of dense suspensions is a fundamental challenge, relevant both for industrial applications, including construction, cosmetics, paints, food processing, pharmaceuticals, and oil recovery, as well as for natural processes such as avalanches, mudflows, and soil erosion. These systems span a wide range of scales, from microscopic colloidal suspensions to meter-scale granular materials.

The properties of dense suspensions are known to evolve with time, a phenomenon commonly referred to as aging~\cite{derec2003aging, jacob2019rheological, cipelletti2005slow}. Aging is usually associated with a progressive slowing down of the dynamics, reflecting changes in the structural configuration of the particle network and in the strength of particle contacts. In dense colloidal suspensions, such as soft glassy systems with contactless interactions and density-matched conditions, Brownian motion of the particles allows them to escape from cages formed by their neighbors, leading to a structural relaxation toward lower free-energy configurations~\cite{courtland2002direct, zaccarelli2009colloidal, hunter2012physics}. As the system ages, such rearrangements become increasingly rare and slow. By contrast, in suspensions where particles touch and form irreversible contacts, the overall structure can rapidly stabilize. In this case, no further particle rearrangements occur, but a different type of aging emerges, driven by progressive strengthening of solid-solid contacts~\cite{bonacci2020contact}.

Structural evolution and contact strengthening are also observed in macroscopic granular suspensions~\cite{andreotti2013granular}. External vibrations, for example, can induce aging-like phenomena. Loosely packed suspensions subjected to moderate-amplitude vibrations initially experience rapid compaction, followed by a slow increase in packing density due to gradual reorganization of the particle network and the decreasing probability of accessing more compact configurations~\cite{kiesgen2015vibration}. Remarkably, even small variations in the initial packing density can dramatically influence the stability and mobility of granular suspensions. For example, the onset of flow in submerged granular sediments can be drastically delayed by dilatancy-related effects~\cite{pailha,Mutabaruka2014, clavaud2017revealing}. Moreover, in densely packed suspensions subjected to very gentle vibrations, even without noticeable changes in the overall density, the contact network can evolve: weak contacts are progressively eliminated, strengthening the packing, and reducing the frequency of plastic events~\cite{kabla2004contact}. On the other hand, granular suspensions can also evolve in the absence of external excitation. Experiments with glass beads in water-filled rotating drums reveal that the avalanche angle increases with resting time, indicating a contact-driven aging process where grain contacts are gradually reinforced by surface chemical reactions~\cite{gayvallet2002ageing}.

A useful framework for distinguishing between colloidal and granular suspensions is provided by the Péclet number, which compares the characteristic time scale of particle motion induced by an external driving force, such as gravity or shear, with that arising from thermal fluctuations~\cite{forterre2018physics}. In colloidal suspensions, particle motion is strongly influenced by thermal agitation and interparticle interactions (e.g. van der Waals or electrostatic forces), which can either promote or prevent particle contact~\cite{wagner2021theory}. In contrast, in granular suspensions, thermal motion is negligible compared to particle weight, and particle dynamics are dominated by direct mechanical contacts and hydrodynamic interactions with the surrounding fluid~\cite{guazzelli2018rheology}. Hence, the behavior of dense suspensions at rest can be characterized by the gravitational Péclet number, $Pe_\mathrm{g} = m g d / k_\mathrm{B} T$, where $m = \pi d^3 \Delta \rho / 6$ is the buoyant particle mass, $g$ gravity, $d$ the particle diameter, and $k_\mathrm{B} T$ the thermal energy. Typically, colloidal suspensions correspond to $Pe_\mathrm{g} \ll 1$, where thermal motion dominates, while granular suspensions lie in the opposite limit, $Pe_\mathrm{g} \gg 1$, where particle weight dominates.

Between these two limits of Péclet, an intermediate regime is expected to exist, in which thermal agitation remains significant compared to particle weight~\cite{ikeda2012unified, Billon2023}. A clear example is provided by sedimented piles of micrometer-sized, electrostatically stabilized particles in water, which display both colloidal and granular features. Tilting such sediments from repose to an initial angle triggers avalanche-like flows, similar to those observed in granular materials under gravity. However, contrary to what is expected, the flow does not stop at the athermal angle of repose ($\theta^* \approx 5.8^\circ$ for frictionless granular materials~\cite{PhysRevE.78.011307, clavaud2017revealing}); instead, the free surface continues to relax through a slow creep regime until it fully flattens. This behavior has been attributed to thermal agitation that allows particles near the free surface to overcome the geometric entanglement of the pile and rearrange over time~\cite{berut2019brownian, lagoin2024effects}. In light of phenomena such as the evolution of the particle network in vibrated granular materials and the slow structural relaxation in colloidal suspensions, it is natural to ask how thermally driven flows of suspensions in the intermediate regime of Péclet may be affected by the waiting time at rest prior to flow.

In this article, we address this question by studying the flow dynamics of dense piles of contactless silica microparticles that are left to rest for a defined period before the flow is induced by tilting. Rotating-drum experiments are conducted at two gravitational Péclet numbers (\(Pe_\mathrm{g} \approx 15 \) and \(264\)), set by particle size, and for initial angles above (\(\theta_\mathrm{start}~=\ang{30}\)) and below (\(\theta_\mathrm{start}~=\ang{5}\)) the athermal angle of repose $\theta^*$. We find that longer resting times lead to delayed flow onsets and reduced flow velocities, clear signatures of aging that are more pronounced at the lower Péclet and when the flow is initiated below $\theta^*$. We show that the observed behavior is not due to chemical particle degradation or the formation of irreversible particle contacts. Importantly, the flow dynamics can be fully rejuvenated by redispersing the particles and allowing them to settle again. We consider several possible mechanisms to account for the physical origin of this aging, including changes in packing fraction, evolution of the packing structure, and modifications of the electrostatic interactions between particles.

\section{MATERIALS AND METHODS}

\subsection{Silica Suspensions}
\label{sec:silica}
Aqueous suspensions with a solid content of 5~wt\% of monodisperse silica microparticles (density $\rho_\mathrm{p}=\SI{1.85}{\gram\per\cubic\centi\meter}$) were obtained from microParticles GmbH. Three systems are defined for the study. Suspension~I and suspension~III are monodisperse, with mean particle diameters $d$ of $1.93\pm0.05~$\si{\micro\meter} and $3.97\pm0.12~$\si{\micro\meter}, respectively. Suspension~II is polydisperse and is prepared by mixing suspensions with $d$ of $1.93\pm0.05~$\si{\micro\meter}, $2.12\pm0.06~$\si{\micro\meter}, and $2.40\pm0.04~$\si{\micro\meter}, in proportions of 25:50:25 by volume.

\subsection{Rotating-drum experiments}
\label{sec:drum}
\begin{figure}[ht]
    \centering
    \includegraphics[width=\columnwidth]{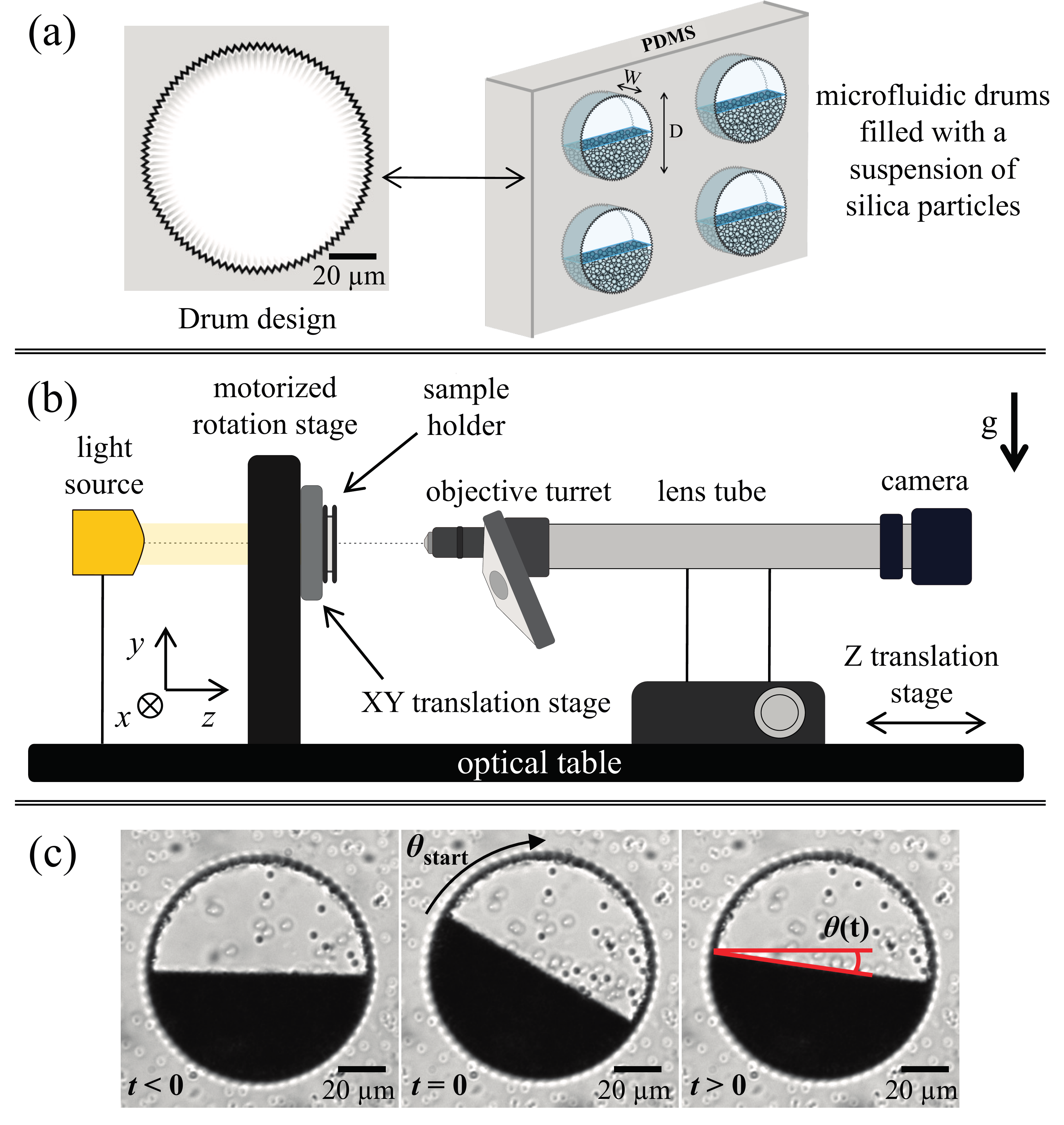} 
    \caption{Schematic representation of the rotating-drum experiments. (a) Drum design with inner walls roughness ($\approx \SI{5}{\micro\meter}$), molded in PDMS to hold a suspension of silica microparticles in a vertical position. (b) Horizontal video-microscopy setup for flow observation at the drum-scale. (c) Characterization of flow dynamics by tilting a pile to an initial angle $\theta_\mathrm{start}$ and monitoring the relaxation of the angle $\theta$ over time $t$.}
    \label{fig:FigI}
\end{figure}

Rotating-drum experiments are conducted following the procedure described in \cite{berut2019brownian, lagoin2024effects}, which we briefly summarize here. Figure~\ref{fig:FigI} illustrates the general workflow: (a) Microfluidic drums (diameter $D = \SI{100}{\micro\meter}$, depth $W = \SI{50}{\micro\meter}$) molded in polydimethylsiloxane (PDMS) are filled with the silica suspensions and sealed with a glass cover slide. (b) The entire device is then vertically placed on a rotation stage coupled to a horizontal microscope, where the suspension is allowed to settle under the influence of gravity. (c) Once a well-defined horizontal pile has formed at the bottom of the drums, the flow is induced by tilting the drums to an initial angle $\theta_{\mathrm{start}}$. This single tilting step is performed at the maximum angular speed of the rotation stage (\SI{720}{\degree\per\second}) to rapidly reach $\theta_{\mathrm{start}}$ and avoid any significant flow before the start of the measurement. The pile relaxation is then tracked by measuring the angle $\theta$ as a function of time $t$. Details regarding PDMS preparation, filling and sealing procedure, and optical setup are provided in the \hyperref[SupMat]{Supplementary Material}.

\subsection{Experimental Procedure}
\label{sec:procedure}
\subsubsection{Pile preparation}
In all cases, the confined suspension in the drums is first resuspended and homogenized by rotating the sample at \SI{90}{\degree\per\second} for 1 hour. After homogenization, the particles are simply allowed to settle at the bottom of the drums at rest for a time $t_\mathrm{w}$, forming a well-defined flat interface with the supernatant fluid.
The sample is then tilted to an initial angle $\theta_{\mathrm{start}}$ of either \SI{5}{\degree} or \SI{30}{\degree}, and 2000 images are acquired at a logarithmic frame rate over a period of 5 hours. This duration corresponds to the characteristic relaxation time observed for piles made of \SI{1.93}{\micro\meter} particles, \textit{i.e.}, the time required for the pile to flatten completely.
For our measurements, we set the reference time $t=0$ as the moment when the sample is tilted.
The waiting time \(t_\mathrm{w}\), which is the time interval between the end of the homogenization step and the moment the sample is tilted, is the control parameter for this study and ranges from 8 minutes to 72 hours.

\subsubsection{Image Analysis and Data Reliability}
A total of 20 drums are recorded simultaneously using a 10$\times$ objective, yielding a pixel resolution of 0.30~\textmu m. Image analysis is performed by contrast differentiation using a contour detection algorithm to automatically identify and track the top surface of the piles. The mean angle $\theta$ is then calculated as a function of time $t$, and averaged across the 20 drums. Each experiment is repeated at least twice to ensure the reliability of the data.

\subsubsection{Additional Experiments: Pile formation and Structure}
Additional images are acquired at different stages of the rotating-drum experiments using a 40$\times$ objective, yielding a pixel resolution of 0.07~\textmu m. Contrast differentiation techniques are applied to measure the height of the pile over time and to extract qualitative features related to the organization of the particles within the pile.

\subsection{Contactless interactions of colloidal silica particles}

The silica particles used in this study are suspended in deionized water and stabilized by partial dissociation of surface silanol groups ($\equiv$SiOH $\rightleftharpoons$ $\equiv$SiO$^{-}$ + H$^+$)~\cite{vigil1994interactions, iler1979chemistry, mcnamee2015time}, which generates an electrostatic repulsive force between them. Solid contact between particles is prevented as long as the effective pressure within the sedimented pile \(P_\mathrm{p}\) remains below a threshold \(P^*\). The critical pressure \(P^*\) is estimated from the electrostatic repulsion force measured by atomic force microscopy (AFM) for the same type of silica particles~\cite{lagoin2024effects}, while \(P_\mathrm{p}\) is determined from the buoyant weight of the particle pile in the microfluidic drums. Under our experimental conditions, the maximum \(P_\mathrm{p}\) remains well below \(P^*\), confirming that the particles never touch each other. Such contactless interactions have been experimentally observed under similar conditions for silica particles with diameters up to \SI{20}{\micro\meter}~\cite{clavaud2017revealing, clavaud2020darcytron}. Full details and theoretical estimates of \(P^*\) and \(P_\mathrm{p}\) are provided in the \hyperref[SupMat]{Supplementary Material}.

\section{Results}

In Subsections \ref{sec:lowPelowtheta} to \ref{sec:highPe}, we characterize the free-surface flow dynamics of the dense piles and their dependence on the waiting time $t_\mathrm{w}$, varying the initial tilting angle $\theta_{\mathrm{start}}$ and the gravitational Péclet number $Pe_\mathrm{g}$. In Subsection \ref{sec:mechanisms} we explore the underlying mechanisms behind the observed behavior.

\subsection{Flow Dynamics in the Thermally Driven Regime at Low Péclet Number}
\label{sec:lowPelowtheta}

\begin{figure}[ht!]
    \centering
    \includegraphics[width=\columnwidth]{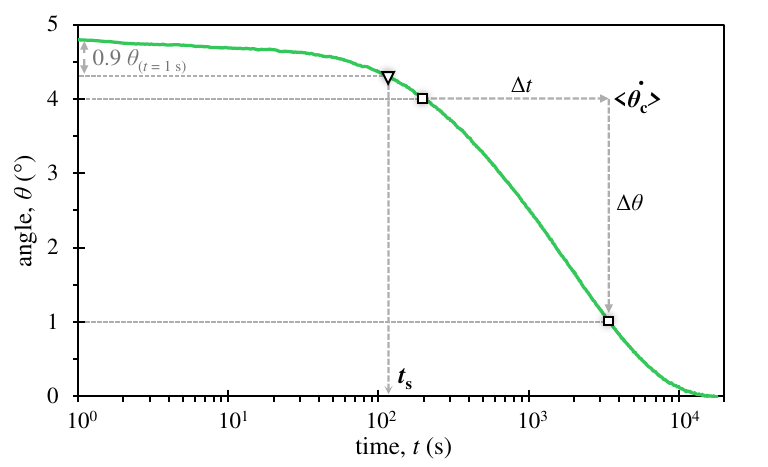}
    \caption{Flow dynamics of gravitationally sedimented piles with low \(Pe_\mathrm{g}\), after being tilted to \(\theta_{\mathrm{start}} = \ang{5}\). Two key parameters are defined: the start time \(t_\mathrm{s}\), when the pile reaches 90\% of the initial angle, and the mean creep speed \(\langle \dot{\theta}_\mathrm{c} \rangle\), representing the average rate of angle decrease during the slow creep relaxation.}
    \label{fig:FigII}
\end{figure}

Figure~\ref{fig:FigII} illustrates the typical flow dynamics of piles formed with suspension~I (\(Pe_\mathrm{g} \approx 15\)), after being tilted to \(\theta_\mathrm{start}~=\ang{5}\). For the first $\sim 100$ seconds, we observe an incipient regime, where the particles start to flow. Then, a long-lasting regime emerges, during which the surface angle decreases, displaying a slow creep relaxation. After approximately 5 hours, the free surface becomes completely flat (\(\theta~=\ang{0}\)), indicating complete arrest of any detectable flow. The emergence of such creep motion, occurring at angles below the expected angle of repose for frictionless non-Brownian suspensions ($\theta^* \approx 5.8^\circ$), is a distinctive signature of thermally driven dynamics~\cite{berut2019brownian, lagoin2024effects}.

\begin{figure}[ht!]
    \centering
    \includegraphics[width=\columnwidth]{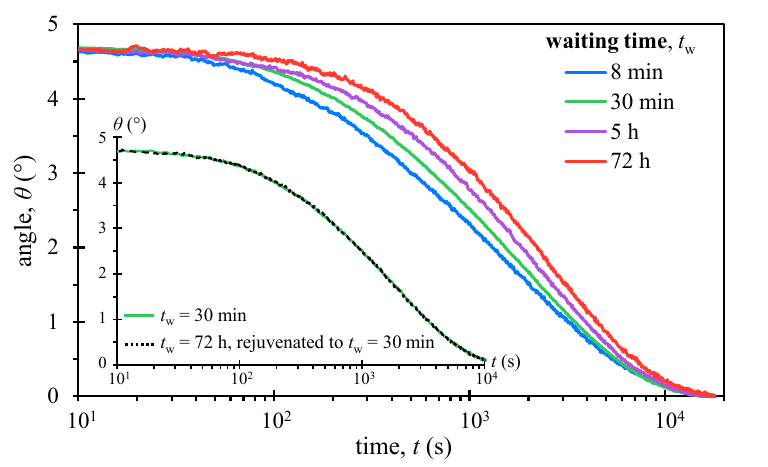} 
    \caption{Effect of the waiting time \(t_\mathrm{w}\) (the time the pile remains at rest before tilting) on the flow dynamics of gravitationally sedimented piles with low \(Pe_\mathrm{g}\), after being tilted to \(\theta_{\mathrm{start}}~=~\ang{5}\). The inset illustrates the rejuvenation of a pile at \(t_\mathrm{w}\)~=~\SI{72}{\hour} after shaking the sample and tilting again at \(t_\mathrm{w}\)~=~\SI{30}{\minute}.}
    \label{fig:FigIII}
\end{figure}

We analyze now how the creep dynamics evolves with the waiting time \( t_\mathrm{w} \). The flow dynamics presented in Figure~\ref{fig:FigIII} reveal that longer waiting times significantly slow down the relaxation of the pile: the longer the pile rests before being tilted, the later the flow begins and the slower the creep proceeds. This behavior suggests the existence of aging in the pile.

To quantitatively characterize our observations, we define two main experimental parameters, schematically introduced in Figure~\ref{fig:FigII}: (i) the start time \(t_\mathrm{s}\), defined as the time required for the pile angle to reach 90\% of the value measured at 1 second, \(\theta(t = \SI{1}{\second})\), and (ii) the mean creep speed \(\langle \dot{\theta}_\mathrm{c} \rangle\), calculated as the average rate of angle decrease during the creep regime (\(\Delta\theta/\Delta t\)). For consistency across experiments, the mean creep speed is calculated between \ang{4} and \ang{1}. By fixing the measurement interval in terms of angles rather than time, we avoid interference from the initial flow delay and capture the dynamics over the range where the angle decays approximately logarithmically with time.

\begin{figure}[ht!]
    \centering
    \includegraphics[width=\columnwidth]{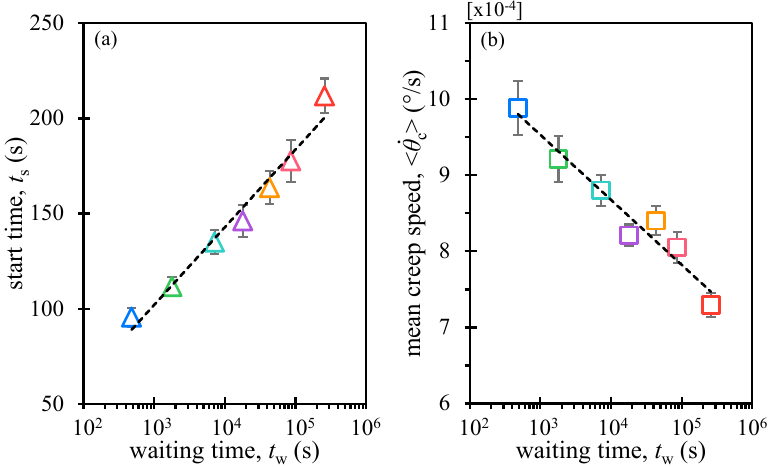} 
    \caption{Characteristic parameters of the creep flow dynamics of gravitationally sedimented piles with low \(Pe_\mathrm{g}\), after being tilted to \(\theta_{\mathrm{start}} = \ang{5}\): (a) start time of the flow \(t_\mathrm{s}\), and (b) mean creep speed \(\langle \dot{\theta}_\mathrm{c} \rangle\), as functions of \(t_\mathrm{w}\). The dashed lines serve as a guide to the eye.
    }
    \label{fig:FigIV}
\end{figure}

Figure~\ref{fig:FigIV} summarizes how the start time \(t_\mathrm{s}\) and the mean creep speed \(\langle \dot{\theta}_\mathrm{c} \rangle\) evolve with the waiting time \( t_\mathrm{w} \). As \( t_\mathrm{w} \) increases, \(t_\mathrm{s}\) increases as well, while \(\langle \dot{\theta}_\mathrm{c} \rangle\) decreases. In particular, both trends show a logarithmic dependence on the waiting time, which supports the interpretation of a slow and progressive aging phenomenon.

Importantly, we observe that the effects of aging on the sedimented piles can be fully rejuvenated. If a pile formed at \(t_\mathrm{w}=t_1\) is re-suspended by the continuous rotation applied during sample homogenization and then tilted at \(t_\mathrm{w}=t_2\), the flow behavior corresponding to the new waiting time $t_2$ is restored. The latter is illustrated in the inset of Figure~\ref{fig:FigIII}, where a pile that had aged for \(t_\mathrm{w}\)~=~\SI{72}{\hour} was rejuvenated to a pile aged for \(t_\mathrm{w}\)~=~\SI{30}{\minute}. Such reversibility indicates that the aging process does not induce a permanent change in the pile, but rather reflects a reversible evolution governed by the time the system remains at rest before triggering the flow.

We further investigated whether piles with low \(Pe_\mathrm{g}\) can undergo aging also during the thermally driven creep flow, where the system is not fully at rest. For this, we prepared piles that were initially aged at rest (\textit{i.e.}, after homogenization and sedimentation) for \(t_\mathrm{w}\)~=~\SI{30}{\minute}, then tilted to \(\theta_{\mathrm{start}} = \ang{5}\) and allowed to fully relax over a period of 5 hours. The resulting flat piles were then directly tilted again to \(\theta_{\mathrm{start}} = \ang{5}\) without resetting the sample by homogenization, and their new relaxation dynamics was measured. We found that the flow behavior and characteristic parameters, \(t_\mathrm{s}\) and \(\langle \dot{\theta}_\mathrm{c} \rangle\), were similar to those of a pile that had aged entirely at rest for \(t_\mathrm{w}\)~=~\SI{5}{\hour}. These observations indicate that pile aging is not interrupted by thermally driven flows; rather, their slow and progressive character may allow the system to continue aging.

\subsection{\label{sec:lowPehightheta}Flow Dynamics in the Gravity Driven Regime at Low Péclet Number}

\begin{figure}[ht!]
    \centering
    \includegraphics[width=\columnwidth]{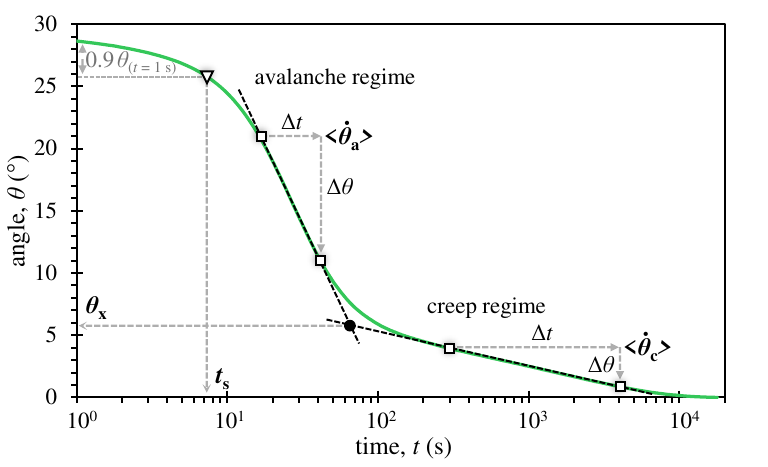} 
    \caption{Flow dynamics of gravitationally sedimented piles with low \(Pe_\mathrm{g}\), after being tilted to \(\theta_{\mathrm{start}} = \ang{30}\). Two flow regimes are defined: a fast avalanche regime and a slow creep regime, characterized by the start time \(t_\mathrm{s}\), the mean avalanche speed \(\langle \dot{\theta}_\mathrm{a} \rangle\), the mean creep speed \(\langle \dot{\theta}_\mathrm{c} \rangle\), and the crossing angle \(\theta_\mathrm{x}\), which marks the transition between them.}
    \label{fig:FigV}
\end{figure}

Changes in pile stability induced by \(t_\mathrm{w}\) were also evaluated when the flow is triggered in the gravity-driven regime, \textit{i.e.}, when the pile is initially tilted above the angle of repose that the material would exhibit in the absence of thermal agitation (\(\theta_{\mathrm{start}} > \theta^* = \ang{5.8}\)). Figure~\ref{fig:FigV} shows the typical flow dynamics of dense piles formed with suspension~I (\(Pe_\mathrm{g}~\approx~15\)), after being tilted to \(\theta_{\mathrm{start}} = \ang{30}\). The dynamics proceeds through two successive regimes: a fast avalanche regime, during which most of the angle decrease occurs, followed by a much slower creep regime that persists until the pile becomes completely flat after approximately 5 hours.

As in the thermally driven regime discussed in Section \ref{sec:lowPelowtheta}, both the start time of the flow \(t_\mathrm{s}\) and the mean creep speed \(\langle \dot{\theta}_\mathrm{c} \rangle\) were calculated, and they are schematically presented in Figure~\ref{fig:FigV}. This procedure was also extended to quantify the mean speed of the avalanche flow \(\langle \dot{\theta}_\mathrm{a} \rangle\) by averaging the angle decrease rate \(\Delta\theta/\Delta t\) along the avalanche regime~\footnote{For comparison across different waiting times, the mean avalanche speed was calculated over the angular range \ang{20}–\ang{10}, capturing the main dynamics of the avalanche regime.}. The transition between the avalanche and creep regimes is identified from the intersection of two logarithmic fits (\(\theta = A \log(t) + B,\) with \(A\), \(B\) constants) applied to the mean portion of each regime. We refer to the pile angle at this intersection as the crossing angle \(\theta_\mathrm{x}\).

\begin{figure}[ht!]
    \centering
    \includegraphics[width=\columnwidth]{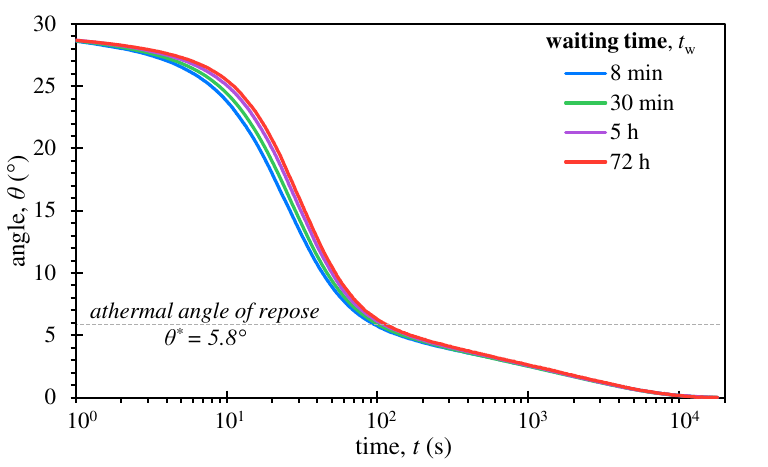} 
    \caption{Effect of the waiting time \(t_\mathrm{w}\) on the flow dynamics of gravitationally sedimented piles with low \(Pe_\mathrm{g}\), after being tilted to \(\theta_{\text{start}} = \ang{30}\).}
    \label{fig:FigVI}
\end{figure}

\begin{figure}[ht!]
    \centering
    \includegraphics[width=\columnwidth]{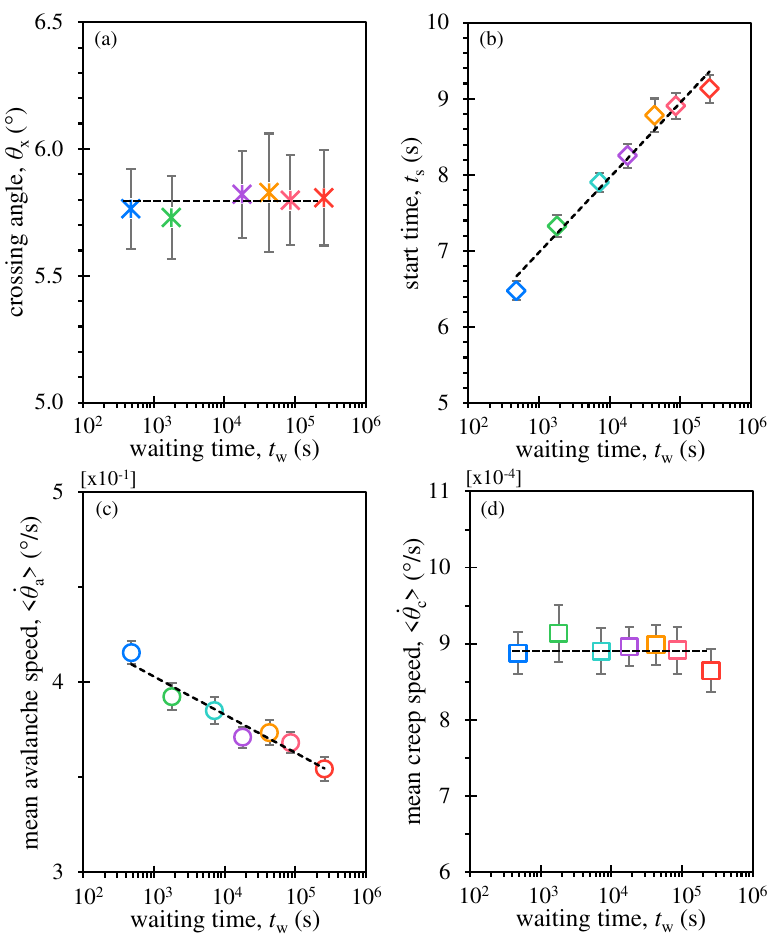} 
    \caption{Characteristic parameters of the flow dynamics of gravitationally sedimented piles with low \(Pe_\mathrm{g}\), after being tilted to \(\theta_{\text{start}} = \ang{30}\): (a) crossing angle \(\theta_\mathrm{x}\), (b) start time of the flow \(t_\mathrm{s}\), (c) mean avalanche speed \(\langle \dot{\theta}_\mathrm{a} \rangle\), and (d) mean creep speed \(\langle \dot{\theta}_\mathrm{c} \rangle\), as functions of \(t_\mathrm{w}\). The dashed lines serve as a guide to the eye.}. 
    \label{fig:FigVII}
\end{figure}

The effects of the waiting time \(t_\mathrm{w}\) on the flow dynamics at low Péclet number and \(\theta_{\mathrm{start}} = \ang{30}\) are shown in Figure~\ref{fig:FigVI}, and the extracted characteristic parameters \(\theta_\mathrm{x}\), \(t_\mathrm{s}\), \(\langle \dot{\theta}_\mathrm{a} \rangle\), and \(\langle \dot{\theta}_\mathrm{c} \rangle\) are summarized in Figure~\ref{fig:FigVII}. 

We find that the crossing angle \(\theta_\mathrm{x}\) is narrowly distributed around \SI[separate-uncertainty = true]{5.8(0.2)}{\degree}, independently of the waiting time \(t_\mathrm{w}\) (Fig.~\ref{fig:FigVII}a). This value coincides with the athermal angle of repose \(\theta^*\) of frictionless granular beds reported in numerical simulations performed in the absence of thermal agitation~\cite{PhysRevE.78.011307}, as well as in experiments with large silica particles in water~\cite{clavaud2017revealing}. These observations indicate that, for piles with low \(Pe_\mathrm{g}\), the crossover from avalanche to creep occurs close to the angle at which flow would arrest in an athermal system (\(\theta_\mathrm{x}\) = \(\theta^*\)), and it is unaffected by \(t_\mathrm{w}\).

In the avalanche regime (\(\theta > \theta^*\)), the waiting time \(t_\mathrm{w}\) moderately influences the flow dynamics of the pile. Longer waiting times lead to an increase in the start time \(t_\mathrm{s}\) and a decrease in the mean avalanche speed \(\langle \dot{\theta}_\mathrm{a} \rangle\) (Fig.~\ref{fig:FigVII}b,c), both of which exhibit a logarithmic dependence on \(t_\mathrm{w}\). These aging-like effects are similar to the trends previously observed when flows are initiated within the creep regime (\(\theta_\mathrm{start} < \theta^*\)). 

In contrast, in the post-avalanche creep regime \(\mbox{(\(\theta < \theta^*\))}\), we found that the influence of \(t_\mathrm{w}\) on the flow behavior disappears. Specifically, the flow profiles converge in the early stages, and the mean creep speed \(\langle \dot{\theta}_\mathrm{c} \rangle\) remains invariant (Fig.~\ref{fig:FigVII}d). These observations suggest that the shear experienced by the pile during the avalanche regime is sufficient to erase the memory of the initial aging, leading the system to a common, history-independent state. As an additional remark, the creep speed of these gravitationally pre-sheared piles is comparable to that of piles aged at rest for a short time (\(t_\mathrm{w}~\approx~\SI{30}{\minute}\)) and tilted to \(\theta_{\mathrm{start}} = \ang{5}\) (Fig~\ref{fig:FigIV}b). This suggests that the gravity driven flow is somehow equivalent to the homogenization procedure that we use to reset the pile.

Taken together, our results for low Péclet number highlight how the effects of aging on the flow dynamics depend on whether the flow is driven by gravity or thermal motion. For purely thermally driven flows, aging remains relevant throughout the entire relaxation process and even continues to evolve during the flow itself (Fig.~\ref{fig:FigIII} and~\ref{fig:FigIV}). However, in gravity-driven flows, the effect is only visible at the beginning of the flow and gradually vanishes as the flow proceeds (Fig.~\ref{fig:FigVI} and \ref{fig:FigVII}).

\subsection{\label{sec:highPe}Flow Dynamics at High Péclet Number}

\begin{figure}[ht!]
    \centering
    \includegraphics[width=\columnwidth]{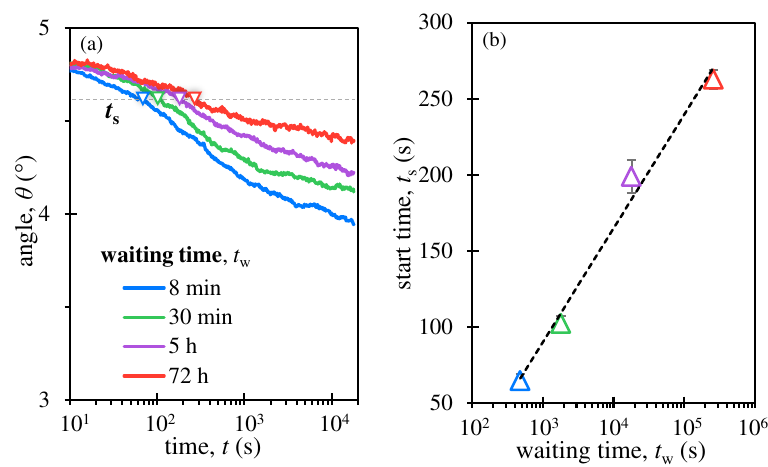} 
    \caption{Effect of the waiting time \(t_\mathrm{w}\) on the creep flow dynamics of gravitationally sedimented piles with high \(Pe_\mathrm{g}\), after being tilted to \(\theta_{\text{start}} = \ang{5}\). \textit{Left}: $\theta$ as a function of time for different \(t_\mathrm{w}\). \textit{Right}: start time of the flow \(t_\mathrm{s}\) (considered here as the time when the pile reaches 95\% of the angle measured during the first second) as a function of \(t_\mathrm{w}\). The dashed lines serve as a guide to the eye.}
    \label{fig:FigVIII}
\end{figure}

We finally investigated whether piles with a higher Péclet number (\(Pe_\mathrm{g} \approx 264\)) undergo aging. Figure~\ref{fig:FigVIII}a shows the flow dynamics of piles formed with suspension~III and tilted to \(\theta_{\mathrm{start}}~=~\ang{5}\). The first observation is that flow still occurs at angles below the angle of repose expected for athermal systems (\(\theta^* \approx \ang{5.8}\)), but it proceeds much slower than in the case of lower \(Pe_\mathrm{g}\). The higher weight of the particles compared to their thermal agitation results in a significantly longer time for the pile to relax~\cite{berut2019brownian}.

The impact of the waiting time \(t_\mathrm{w}\) on these thermally driven flows manifests as a slower initial flow and a delay that persists, and even becomes more pronounced, during the creep flow. As observed at low $Pe_\mathrm{g}$, the start time $t_\mathrm{s}$ increases logarithmically with \(t_\mathrm{w}\) (Fig.~\ref{fig:FigVIII}b).

\begin{figure}[ht!]
    \centering
    \includegraphics[width=\columnwidth]{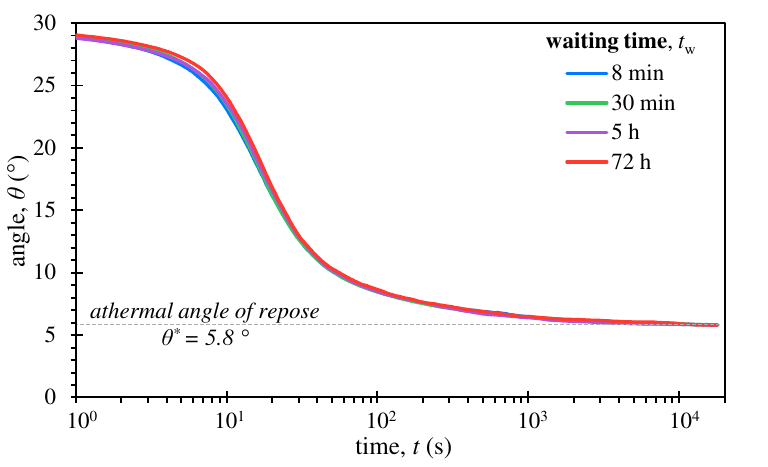} 
    \caption{Effect of the waiting time \(t_\mathrm{w}\) on the flow dynamics of gravitationally sedimented piles with high \(Pe_\mathrm{g}\), after being tilted to \(\theta_{\text{start}} = \ang{30}\).}
    \label{fig:FigIX}
\end{figure}

\begin{figure}[ht!]
    \centering
    \includegraphics[width=\columnwidth]{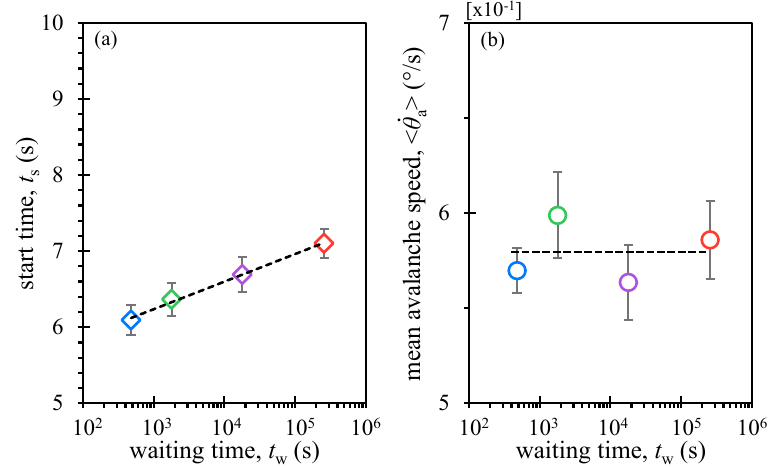} 
    \caption{Characteristic parameters of the flow dynamics of gravitationally sedimented piles with high \(Pe_\mathrm{g}\), after being tilted to \(\theta_{\text{start}} = \ang{30}\): (a) start time of the flow \(t_\mathrm{s}\), (b) mean avalanche speed \(\langle \dot{\theta}_\mathrm{a} \rangle\), as functions of \(t_\mathrm{w}\). The dashed lines serve as a guide to the eye.}
    \label{fig:FigX}
\end{figure}

Consistently, the role of gravity in erasing the effects of aging becomes evident when the piles are initially tilted to an angle above the athermal angle of repose (\(\theta_{\mathrm{start}}~=~\ang{30}\)), as illustrated in Figure~\ref{fig:FigIX}. The influence of \(t_\mathrm{w}\) appears only as a slight increase in the start time (Fig.~\ref{fig:FigX}a), but is quickly erased by the avalanche itself. The flow profiles rapidly converge in less than $\sim 100$ seconds, and the mean avalanche speed \(\langle \dot{\theta}_\mathrm{a} \rangle\) remains essentially unchanged with \(t_\mathrm{w}\) in this case (Fig.~\ref{fig:FigX}b).

Note that, after the avalanche, the creep regime proceeds extremely slowly and appears to converge around the athermal angle of repose \(\theta^* \approx \ang{5.8}\), as shown in Figure~\ref{fig:FigIX}. This suggests a transition from avalanche to creep similar to that observed at low \(Pe_\mathrm{g}\), without any effect of \(t_\mathrm{w}\). However, at high \(Pe_\mathrm{g}\), the creep is too slow to fully develop within the time-frame of the experiment, and therefore neither the crossing angle \(\theta_\mathrm{x}\) nor the mean creep speed \(\langle \dot{\theta}_\mathrm{c} \rangle\) can be measured\footnote{At high \(Pe_\mathrm{g}\), we expect that after the avalanche regime a creep flow would develop and persist below the athermal angle of repose \(\theta^*\). This idea is supported by the existence of a creep regime when the suspension is tilted from \(\theta_\mathrm{start} = \SI{5}{\degree}\) (see Fig.~\ref{fig:FigVIII}). However, the full characterization of this regime would require timescales that are too long to be accessed in the present experiments.}.

Taken together, our observations at low and high Péclet numbers show that sedimented piles exhibit aging while at rest prior to flow, manifesting itself as a logarithmic increase in characteristic parameters such as start time and mean flow speed. These effects are clearly stronger when the flow is thermally driven, and remain visible even more than one hour after flow initiation. Interestingly, when the flow is driven by gravity (\textit{i.e.}, initiated above the athermal angle of repose), the aging-induced delays tend to vanish over time, making the differences between flow curves with different waiting times progressively less pronounced as the flow proceeds.

\subsection{\label{sec:mechanisms}Mechanisms Underlying Aging}

Several mechanisms may contribute to the aging of sedimented piles of micrometer-sized, electrostatically stabilized silica particles in water. In the following subsections, we examine the roles of pile compaction, structural evolution, and surface-chemistry effects in the samples.

\subsubsection{Changes in packing fraction}

First, the initial packing fraction of our sedimented piles is established by the combined effects of gravity, electrostatic repulsion, hydrodynamic interactions, and thermal agitation, which together determine how the particles settle and organize~\cite{forterre2018physics, wagner2021theory, guazzelli2018rheology}. Although the absence of frictional contacts between particles is expected to stabilize the particle network near its maximum accessible random close-packing fraction~\cite{clavaud2017revealing, PhysRevE.78.011307}, it is not a priori clear to what extent this state is actually reached during sedimentation~\cite{jerkins2008onset}. Any residual evolution of the packing fraction during the waiting time \(t_\mathrm{w}\) could, in principle, enhance the stability of the pile by allowing the particles to adopt more compact local configurations, potentially facilitated by thermal fluctuations. This scenario would be somehow reminiscent of the effect of mechanical vibrations in frictional granular media, which enable particles to rearrange into denser, more mechanically stable configurations~\cite{kiesgen2015vibration, kabla2004contact, pailha}. 

\begin{figure}[ht!]
    \centering
    \includegraphics[width=\columnwidth]{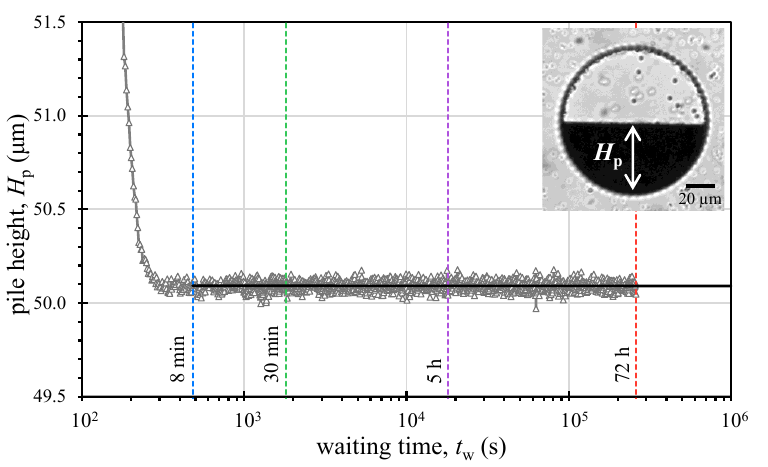} 
    \caption{Long-term sedimentation profile of particles with a diameter of $1.93 \pm 0.05~$\si{\micro\meter}. The pile height \(H_\mathrm{p}\) is defined as the mean vertical position of the sedimentation front within the microfluidic drum (see inset). The horizontal black line indicates the steady-state pile height $\overline{H_\mathrm{p}}$.}
    \label{fig:FigXI}
\end{figure}

To assess whether our colloidal sediments undergo measurable changes in their packing fraction during the waiting time, we tracked the height of the bed over time. Figure~\ref{fig:FigXI} shows the long-term sedimentation profile of particles with a diameter of $1.93 \pm 0.05~$\si{\micro\meter} inside the drums. Here, $H_\mathrm{p}$ denotes the mean vertical position of the sedimentation front measured from the bottom of the container. During gravitational settling (only shown for $t_\mathrm{w} > 100$ seconds), $H_\mathrm{p}$ decreases monotonically until a well-defined pile forms at the bottom of the drum in less than \SI{8}{\minute}.  From this point onward, the height of the sediment ceases to decrease and instead fluctuates around a well-defined steady-state value $\overline{H_\mathrm{p}}$. Similar behavior is observed for piles formed with larger particles (suspension~III, $d = 3.97 \pm 0.12~$\si{\micro\meter}), which settle and stabilize even more rapidly.

These observations suggest that no measurable compaction occurs after that naturally associated with the advance of the sedimentation front during the formation of the pile. Based on the pixel resolution of our image analysis ($\delta H_\mathrm{p} \simeq \SI{0.1}{\micro\meter}$), we estimate that any change in the overall packing fraction is constrained to be below $\sim 0.2\%$ (from $|\Delta \phi|/\phi \lesssim \delta H_\mathrm{p} / \overline{H_\mathrm{p}}$, with $\overline{H_\mathrm{p}} \simeq \SI{50}{\micro\meter}$). Therefore, as we do not observe any long-term effect, compaction by itself is insufficient to explain the aging manifested by the sedimented piles at rest.

\subsubsection{Structural evolution}

\begin{figure}[ht!]
    \centering
    \includegraphics[width=\columnwidth]{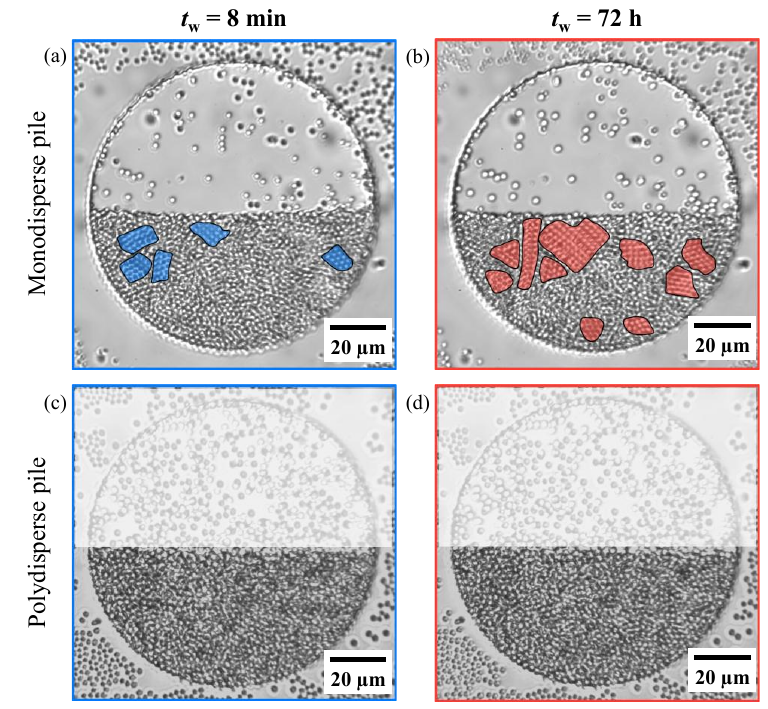} 
    \caption{Microphotographs of monodisperse (a-b) and polydisperse (c-d) piles: typical structure of the first lateral layer of particles within the pile at \(t_\mathrm{w}\)~=~\SI{8}{\minute} and \(t_\mathrm{w}\)~=~\SI{72}{\hour}. Regions of local particle organization are highlighted in color.}
    \label{fig:FigXII}
\end{figure}

Second, systems of monodisperse particles are known to evolve from amorphous to crystalline configurations when subjected to dynamical fluctuations. In macroscopic granular materials, such fluctuations are typically injected by mechanical vibrations that promote particle rearrangements toward more ordered packings~\cite{pouliquen1997crystallization, vanel1997rise, nicolas2000compaction, kabla2004contact}. In dense colloidal suspensions, this evolution occurs spontaneously and is driven by microscopic fluctuations arising from the thermal agitation of the solvent, which allows the particles to rearrange and minimize free energy~\cite{gasser2009crystallization,pusey1987observation}, even under quiescent conditions or during slow gravitational sedimentation~\cite{pusey1986phase, davis1991settling}. In our experiments, we worked with highly monodisperse suspensions, in which such dynamics may in principle occur. Figure~\ref{fig:FigXII}a,b shows the first lateral layer of particles (from the glass wall inward) of typical piles prepared from suspension~I (\(Pe_\mathrm{g} \approx 15\)) at \(t_\mathrm{w}\)~=~\SI{8}{\minute} and \(t_\mathrm{w}\)~=~\SI{72}{\hour} \footnote{These experiments were carried out in \SI{10}{\micro\meter}-deep devices. While the relaxation of the piles when tilted is slower in these thinner devices, the same $t_\mathrm{w}$-dependence of the flow dynamics is observed as in the thicker devices (\SI{50}{\micro\meter}-deep).}. We note that regions of local order appear immediately after sedimentation. Longer waiting times seem to promote both the growth of existing ordered domains and the emergence of new ones. However, it is not possible to establish unambiguously whether these local regions correspond to actual crystallization that extends into the depth of the drums. Consequently, on the basis of these observations, it remains unclear to what extent the aging manifested by monodisperse piles depends on structural ordering and its temporal evolution.

\begin{figure}[ht!]
    \centering
    \includegraphics[width=\columnwidth]{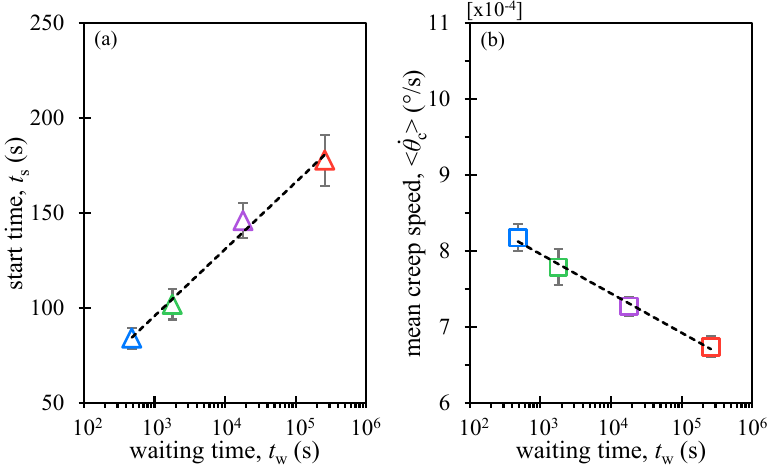} 
    \caption{Effect of the waiting time \(t_\mathrm{w}\) on (a) the start time of the flow \(t_\mathrm{s}\) and (b) the mean creep speed \(\langle \dot{\theta}_\mathrm{c} \rangle\) of gravitationally sedimented piles prepared from suspension II (polydisperse), after being tilted to \(\theta_{\text{start}} = \ang{5}\). The dashed lines serve as a guide to the eye.}
    \label{fig:FigXIII}
\end{figure}

Therefore, to address this question, we turned to a system in which crystallization is strongly suppressed by introducing polydispersity~\cite{zaccarelli2009crystallization}. We prepared sedimented piles using suspension~II, a mixture of particles of different sizes, and probed their flow response as a function of the waiting time \(t_\mathrm{w}\). Visualizations of the first particle layer in the drums (Fig.~\ref{fig:FigXII}c,d) revealed no signs of local ordering or crystal growth over time. However, when tilted to \(\theta_{\text{start}} = \ang{5}\), these polydisperse piles exhibited the same aging signatures as their monodisperse counterparts. Figure~\ref{fig:FigXIII} reports the start time of the flow \(t_\mathrm{s}\) and the mean creep speed \(\langle \dot{\theta}_\mathrm{c} \rangle\), both of which display a logarithmic dependence on $t_\mathrm{w}$. These results demonstrate that the aging in the flow response of our sedimented piles is not a mere reflection of crystallization or long-range structural ordering. If the underlying mechanism is structural in nature, it cannot be trivially attributed to the tendency of the system toward ordering.  

We conclude this section by emphasizing that the aging manifested in our colloidal suspensions is reminiscent of the mechanical aging broadly reported in colloidal glasses. In these systems, rheological measurements show a clear age-dependent dynamics: the material’s resistance to flow or deformation increases with waiting time, but in a progressively slower way~\cite{derec2003aging, jacob2019rheological, cipelletti2005slow}. Notably, while this evolution has been shown to occur without measurable changes in long-range structural order~\cite{cianci2006invariance}, it has been consistently linked to slow, cooperative particle rearrangements driven by thermal fluctuations, which in turn become increasingly rare and slow as the system ages~\cite{courtland2002direct, zaccarelli2009colloidal, hunter2012physics}. It is important to note, however, that colloidal glasses are typically studied in bulk suspensions, where particles interact as quasi-hard spheres and sedimentation is suppressed by density matching. In contrast, in our system, particles settle under gravity to form a sediment, experience short-range electrostatic repulsion, and the aging signatures we observe emerge at the free surface of the particle bed.  

Despite the differences, we hypothesize that in our sediments, thermal fluctuations, still relevant compared to particle weight, could allow particles in the upper layers near the free surface to undergo slow cooperative rearrangements during the waiting time. Such rearrangements may bring the packing into energetically more favorable configurations without inducing measurable changes in the packing fraction or implying long-term structural order. The enhanced stability acquired at rest would then manifest as a higher resistance to leaving these locally favorable configurations when the pile is tilted, providing a natural explanation for the observed delayed onset and slower progression of flow.  

\subsubsection{Particle surface chemistry}

Finally, dynamics governed by surface chemistry could also contribute to the evolving stability of the piles. 

The formation of structured molecular layers on silica surfaces in water is still a matter of debate, but there is evidence pointing to the existence of a gel-like layer a few nanometers thick~\cite{vigil1994interactions, bitter2013anomalous}. These layers confer complex behavior to silica particles: they can reduce the effect of short-range van der Waals attractions through an additional steric repulsion, but, when forced into contact, adhesive forces may emerge between them, strengthening interparticle bonds over time~\cite{mcnamee2015time}. This phenomenon has been invoked to explain the aging observed in frictional granular systems across different length scales~\cite{bonacci2020contact, gayvallet2002ageing}. 

However, in our system, interparticle separations are expected to remain on the order of a few times the Debye length (\(\lambda_\mathrm{D} \approx 65\ \mathrm{nm}\); see \hyperref[SupMat]{Supplementary Material}) due to the low effective pressure in the particle pile, well beyond the range of gel-like layers. AFM measurements further indicate that there are no adhesive interactions or attractive forces that could promote particle aggregation~\cite{lagoin2024effects}. Moreover, the aging we observe does not depend on the age of the suspension or the sample (once in the drum), but on the time elapsed after the pile forms, as demonstrated by the reproducibility of behaviors across different samples (up to one year apart) and repeated measurements on the same sample (up to one month apart). Therefore, irreversible processes such as the formation and growth of gel-like layers on silica surfaces are not expected to significantly affect particle behavior on this timescale.

Even though our particles always remain a few hundreds of \si{\nano\meter} apart, and therefore never touch each others, direct contact between particle surfaces is not a prerequisite for other surface-level dynamics to emerge. For example, charge regulation, in which the surface charge density of particles dynamically adjusts to the local ionic environment, can change the particles interactions over time~\cite{israelachvili2011intermolecular, bakhshandeh2019charge, bakhshandeh2020charge}. Silica suspensions prepared with ultrapure water inherently contain ions, arising from the partial and reversible dissociation of surface silanol groups, which release protons (H$^+$)~\cite{vigil1994interactions,iler1979chemistry}. Theoretical studies have shown that, despite the free diffusion of protons in bulk solution, their mobility can become limited in complex environments, such as densely packed assemblies, confined geometries, or near charged interfaces, leading to slow and time-dependent processes~\cite{ritt2022thermodynamics, curk2021charge}. Experimentally, AFM measurements on silica surfaces submerged in water have shown that the electrostatic repulsive force between particles decreases with resting time, with changes still visible after 60 minutes of rest. These observations are consistent with a slow redistribution of ions between the bulk solution and the particle surfaces~\cite{vermeulen1995slow, mcnamee2015time}.  

In the case of our silica particles, their electrostatic equilibrium in bulk suspension (when dispersed inside the drums) may differ from that within a dense pile (when sedimented). The close proximity of neighboring particles can perturb the local ionic distribution, leading to adjustments in the surface charge density. The adaptation of the particles to this altered electrostatic environment could introduce an additional timescale associated with proton diffusion and partial surface neutralization, a process that may be further slowed by the presence of a gel-like layer. Consequently, a gradual reduction of the electrostatic repulsion between particles resulting from this charge-regulation dynamics would broaden the accessible energy landscape, thereby facilitating thermally driven particle rearrangements. This additional contribution from surface chemistry to the enhanced stability of the pile appears consistent with the rejuvenating nature of the observed aging: once the particles are re-dispersed back into bulk suspension, the electrostatic conditions are effectively restored, and the waiting time is reset to zero.

\section{CONCLUSION}
\label{sec:conclusion}

We have shown that sedimented piles of micrometer-sized, electrostatically stabilized silica particles undergo aging when left at rest. This effect, revealed through free-surface flows triggered by tilting, is manifested as delayed flow onset and reduced flow velocity, both evolving logarithmically with the resting time. The extent and persistence of these changes are greater when the flow is driven by thermal fluctuations (creep flows below the athermal angle of repose) than when driven by gravity (avalanche flows above it). While avalanche flows tends to erase the signatures of aging, creep flows are slow enough to preserve them. Crucially, flow dynamics was found to be fully rejuvenated by redispersing particles, demonstrating that observed changes do not depend on the age of the sample, but only on the time elapsed between sediment formation and flow initiation. Moreover, we have shown that this aging phenomenon occurs without a measurable evolution of the packing fraction in the pile at rest, and in the absence of crystallization. Finally, the electrostatically stabilized particles never come into contact with one another, which rules out the possibility of chemical strengthening of inter-particle contacts. These findings provide evidence of a form of aging that emerges from the intrinsic thermal activity and electrostatic nature of the system. Although not colloidal in the classical sense, these sediments lie in an intermediate regime between colloidal and granular suspensions, where thermal agitation still play a decisive role in shaping flow dynamics and promoting the emergence of time-dependent effects.

\section*{Data availability}
The data that support the findings of this article are openly available~\cite{DataZenodo}.

\begin{acknowledgments}
The authors acknowledge the support of the French Agence Nationale de la Recherche (ANR), under grant ANR-21-CE30-0005 (JCJC MicroGraM).
\end{acknowledgments}

\bibliography{biblio}

\newpage
$~$
\newpage

\onecolumngrid

\section*{Supplementary Material}
\label{SupMat}

\subsection{Fabrication of PDMS Microfluidic Drums}
PDMS stamps are manufactured using standard soft photolithography techniques \cite{Xia, MacDonald}. Our design consists of an array of $100 \times 100$ drums, surrounded by a manually carved moat that serves as a water reservoir to prevent water evaporation from the sample. The drums have a diameter of $100\,\si{\micro\meter}$ and a depth of \SI{50}{\micro\meter}, the inner walls of which feature a periodic triangular pattern with an opening and a depth of approximately \SI{5}{\micro\meter}. This pattern creates a rough surface that prevents the piles from sliding along the walls during flow. The PDMS stamps are sealed on their underside on a glass slide using plasma treatment at \SI{0.5}{\torr} and \SI{30}{\watt} for \SI{2}{\minute}, and left to rest for \SI{72}{\hour}. The latter ensures that the lipophilic nature of PDMS is recovered prior to use \cite{Bacharouche}.

\subsection{Filling of the Microfluidic Drums with the Silica Suspensions}
The microfluidic drums are filled with the silica suspensions following a specific protocol. Initially, the PDMS sealed on glass device is thoroughly cleaned by rinsing with isopropyl alcohol (IPA, Merck) and distilled water (DW) supplied by an ELGA Purelab\textsuperscript{\textregistered} Flex system (\SI{18.2}{\mega\ohm\cm}). The device is then immersed face down in a cell filled with DW and placed in an ultrasonic bath for 15 minutes to remove any impurities from the drums. After sonification, it is transferred to a Petri dish with the drums facing upward and placed in a vacuum chamber to remove trapped air for 10 minutes. The device is then reimmersed in a fresh DW cell to fill the drums with water.  

After this step, the device is placed on a sample holder with the drums facing upward. The carved moat is filled with DW, and a small droplet (approximately \SI{30}{\micro\liter}) of the silica suspension is carefully deposited on top of the drum array using a micropipette. The particles are allowed to sediment for 2 minutes, after which the excess droplet is removed by gently pressing a glass slide, pre-washed with IPA and DW, onto the PDMS. Finally, the device is secured with a top aluminum plate fastened with screws, which presses the glass slide against the PDMS to ensure proper sealing of the drums. The particles typically occupy 50\% of the drum volume, as depicted in Figure 1(c).

\subsection{Videomicroscopy Setup}
Drum-scale observations are performed using the videomicroscopy setup shown in Figure 1(b). This custom-made horizontal microscope consists of a CCD camera (Basler acA2440-75um) connected to a microscope turret equipped with long working distance objectives (Olympus MPLFLN $\times$10 and LUCPLFLN $\times$40) via a lens tube (InfiniTube\texttrademark\ Standard). The setup also includes a motorized rotation stage (Newport URB100CC), on which the sample holder containing the PDMS device is fixed. The rotation axis of the stage is precisely aligned with the optical axis of the videomicroscopy system, and a manual 2D translation stage allows for mounting and dismounting the sample. To minimize external vibrations, the entire system is mounted on an optical table with passive isolation stands (Thorlabs PWA075).

\subsection{Estimation of the critical pressure for contactless particle interactions}  

The electrostatic repulsive force arising from double-layer interactions between two spherical particles separated by a surface-to-surface distance $r$ is described by  

\begin{equation}
F_{\text{rep}} = \frac{dZ}{4\lambda_\mathrm{D}} \exp \left( -\frac{r}{\lambda_\mathrm{D}} \right),
\label{eq:force}
\end{equation}  

where $d$ is the particle diameter, $Z$ is an interaction constant related to the surface potential, and $\lambda_\mathrm{D}$ is the Debye length~\cite{israelachvili2011intermolecular, kobayashi2005aggregation}. 

In the limiting case where the particles approach solid contact, the minimal separation is set by the typical surface roughness, $l_\mathrm{r}$. Accordingly, the critical pressure for contact, $P^*$, can be expressed as the repulsive force at $r = 2l_\mathrm{r}$, divided by the projected cross-sectional area of a particle, yielding

\begin{equation}
P^* = \frac{F_{\text{rep}}}{\pi d^2 / 4} \Big|_{r = 2 l_\mathrm{r}}.
\label{eq:Pcrit}
\end{equation}  

In a dense particle pile, the possibility of inter-particle contact can be assessed by comparing the critical pressure for contact $P^*$ with the effective pressure within the pile $P_\mathrm{p}(z)$~\footnote{Note that thermal agitation is negligible compared to the electrostatic interaction energy. At room temperature, the thermal energy is $E_{\text{thermal}} = k_\mathrm{B} T \approx 4.1 \times 10^{-21}\,\text{J}$, whereas the repulsive energy between two silica particles of $d = \SI{1.93}{\micro\meter}$ separated by a distance equal to the Debye length ($r = \lambda_\mathrm{D}$) is approximately $E_{\text{rep}} = (dZ / 4) \exp(-1) \approx 3.42 \times 10^{-18}\,\text{J}$, larger by about three orders of magnitude}. The latter is calculated from the buoyant weight of the particles. For a sedimented pile of height $H_\mathrm{p}$, particle volume fraction $\phi$, and particle density $\rho_\mathrm{p}$, with a fluid column of height $H_\mathrm{f}$ and density $\rho_\mathrm{f}$ above it, the hydrostatic pore pressure is $u(z) = \rho_\mathrm{f} g (H_\mathrm{f} + z)$, while the total stress is $\sigma(z) = \rho_\mathrm{f} g H_\mathrm{f} + [\phi \rho_\mathrm{p} + (1-\phi)\rho_\mathrm{f}]gz$. Here, the depth $z$ is measured downward from the free surface of the pile. The effective pressure within the pile is then calculated as the difference between the total stress $\sigma(z)$ and the pore pressure $u(z)$, yielding

\begin{equation}
P_\mathrm{p}(z) = \phi (\rho_\mathrm{p} - \rho_\mathrm{f}) g z,
\label{eq:confining}
\end{equation}
which depends solely on the buoyant weight of the particles and increases linearly with depth.

Using the AFM-measured values from Lagoin et al. (2023)~\cite{lagoin2024effects} for silica particles supplied by the same manufacturer ($Z = 6.23 \times 10^{-12} \,\text{J/m}$, $\lambda_\mathrm{D} = 65.4 \pm 11.2 \,\text{nm}$, and $l_\mathrm{r} \approx 2 \pm 1 \,\text{nm}$), we obtain $P^* \approx 15 \pm 3$ Pa for particles with $d = \SI{1.93}{\micro\meter}$ (suspension I) and $P^* \approx 7 \pm 1$ Pa for particles with $d = \SI{3.97}{\micro\meter}$ (suspension III). Taking $\phi = 0.512$ from 3D confocal reconstructions of sedimented piles of silica particles \SI{2.12}{\micro\meter} in diameter under confining conditions similar to those in our experiments~\cite{lagoin2024effects}, the maximum pressure within the pile (evaluated at $z = \overline{H_\mathrm{p}} \simeq \SI{50}{\micro\meter}$) is estimated as $P_{\mathrm{p,max}} \approx \SI{0.2}{\pascal}$. Given that $P_\mathrm{p}$ is nearly two orders of magnitude smaller than $P^*$ across all suspensions, the particles in the pile are expected to remain well separated, with interparticle distances far exceeding their surface roughness. This supports the assumption of a contactless configuration in our experiments. We further estimated the mean interparticle distance by equating the electrostatic pressure (from Eq.~\ref{eq:force}) to the maximum pressure within the pile ($P_{\mathrm{p,max}}$), which yields separations of at least three to four times the Debye length ($r \gtrsim 3$–$4\,\lambda_\mathrm{D}$).

\end{document}